\documentclass[conference]{IEEEtran}
\usepackage[letterpaper, left=1in, right=1in, bottom=1in, top=0.75in]{geometry}
\usepackage{cite}
\usepackage{array}	
\usepackage{flushend}	
\usepackage{algorithm}
\usepackage{algorithmic}

\ifCLASSOPTIONcompsoc
  \usepackage[caption=false,font=normalsize,labelfont=sf,textfont=sf]{subfig}
\else
  \usepackage[caption=false,font=footnotesize]{subfig}
\fi

\ifCLASSINFOpdf
   \usepackage[pdftex]{graphicx}
   \DeclareGraphicsExtensions{.pdf,.jpeg,.png}
\else
\fi

\usepackage{amsmath}
\usepackage{amssymb}
\usepackage{amsthm}     
\usepackage{blkarray}   

\usepackage{color}  
\usepackage{url}
\mathchardef\mhyphen="2D 
\newenvironment{eqn}%
{\begin{eqnarray}\begin{aligned}}%
{\end{aligned}\end{eqnarray}}

{\begin{small}\begin{eqnarray}\begin{aligned}}%
{\end{aligned}\end{eqnarray}\end{small}}
\IEEEoverridecommandlockouts
\begin{document}

\title{ANN-Based Detection in MIMO-OFDM Systems with Low-Resolution ADCs}

\author{\IEEEauthorblockN{Shabnam Rezaei and Sofiene Affes}
\vspace{0.3cm}
\IEEEauthorblockA{Institut National de la Recherche Scientifique (INRS-EMT)\\ University of Quebec, Montreal, QC, Canada\\
Emails: \{shabnam.rezaei, affes\}@emt.inrs.ca}\\
\thanks{Work supported by the Discovery Grants and the CREATE PERSWADE (www.create-perswade.ca) Programs of NSERC.}
}

\maketitle

\begin{abstract}
In this paper, we propose a  multi-layer artificial neural network (ANN) that is trained with the Levenberg-Marquardt algorithm for use in signal detection over multiple-input multiple-output orthogonal frequency-division multiplexing (MIMO-OFDM) systems, particularly those with low-resolution analog-to-digital converters (LR-ADCs). We consider a blind detection scheme where data symbol estimation is carried out without knowing the channel state information at the receiver (CSIR)---in contrast to classical algorithms.  The main power of the proposed ANN-based detector (ANND) lies in its versatile use with any modulation scheme, blindly, yet without a change in its structure. We compare by simulations this new receiver with conventional ones, namely, the maximum likelihood (ML), minimum mean square error (MMSE), and zero-forcing (ZF), in terms of symbol error rate (SER) performance. Results suggest that ANND approaches ML at much lower complexity, outperforms ZF over the entire range of assessed  signal-to-noise ratio (SNR) values, and so does it also, though, with the MMSE over different SNR ranges.

\end{abstract}

\IEEEpeerreviewmaketitle

\section{Introduction}
Most MIMO detection techniques developed so far requir are based on having perfect CSIR to perform detection. Therefore, channel estimation for such techniques is inevitable \cite{far2018}, \cite{luo2018channel}. In fact, the more accurate is the CSIR, the better would be the detection performance. The most popular channel estimation algorithms are ML, least-square (LS), MMSE. \cite{ophysical2017}  uses an LS algorithm to perform MIMO channel estimation.  \cite{jnear2016} considers performing ML-based channel estimation to predict the CSIR. The drawback of such techniques is that they need to perform computationally-complex and time-consuming matrix manipulations \cite{SEYMAN2013estimation},\cite{Dorner2018deep}. Hence, such classical algorithms are suitable for small-scale (i.e.,the few antennas) CSIR estimation in wireless communication systems. On the other hand, future wireless systems like mmWave systems are aimed to provide communication links with high data rates by using wide bandwidths. As the system bandwidth increases, the sampling rate of an analog-to-digital converter (ADC) should linearly increase too. Unfortunately, higher sampling rates lead to larger power consumption of ADCs. Using LR-ADCs has been considered as a cost-effective solution to reduce power consumption of systems requiring high-speed ADCs \cite{r1999adc}. By deploying LR-ADCs in a system, the accurate CSIR cannot be obtained from pilot signals because of the excess error in quantization. To solve this problem, several channel estimation methods have been developed. One common problem in all of such algorithms is that they require a huge amount of pilot signals to overcome the excess error. 

Blind detection in MIMO communication systems have been studied extensively. For instance, a completely-blind MIMO detection technique has been developed in \cite{langblind1994}, and the K-means clustering algorithm was proposed in \cite{liangcluster2013} for blind detection. The main drawback of such algorithms for blind detection is that they have  high implementation complexity in order to produce accurate CSIR, which may not be affordable in practical communication systems. Further, semi-blind MIMO detection methods, which perform data detection and channel estimation jointly, have been developed and shown to outperform coherent detection methods. However, the key problem in related to these blind techniques is that they are only available for some particular types of modulations like space-shift-keying or phase-shift-keying (PSK).

Recently, many researchers started looking for innovative solution based on machine learning (ML). For example, for the main issue of blind detection discussed above the authors in \cite{Jeon2017Blind} have rethought the detection operation from scratch without requiring CSIR by applying ML algorithms. They have used a K-nearest neighbors (KNN) classifier for symbol detection over MIMO systems. Therein, it is shown that KNN is comparable in performance to the conventional detection algorithms. However, its accuracy may decrease in multi-user MIMO systems because of its inherent limitations compared to other learning algorithms like ANN. 

In this paper, we propose a blind signal detection approach based on ANN for MIMO-OFDM systems with perfect ADCs and with LR-ADCs when explicit CSIR is not known to the receiver. Neural networks have offered the state-of-the-art solutions in many domains other than classifications. Since we can view the detection as an act of classification, we choose this technique as a classifier. Training a neural network involves many hyper-parameters controlling the size and structure of the network and the optimization procedure which can help achieve better detection over large-scale systems. The proposed approach consists of two phases. In the first one, the transmitter sends a sequence of data symbols so that the receiver learns a nonlinear function that describes input-output relations of the system. In the second phase, using the trained neural network, the receiver detects the data symbols. This new approach can be regarded as a classification problem in supervised learning. The classifier maps the received signal to one of the possible symbol vectors.The main advantage of our ANN-based detector is that its operation is independent of the modulation scheme adopted by the system---contrary to conventional detection techniques where the receiver structure should be adapted to the modulation scheme.
The rest of the paper is arranged as follows. In Section \ref{sec:SystemModel}, the MIMO-OFDM system model is explained. Section \ref{sec:MLP} describes the multi-layered percepteron neural network and the training algorithm. In Section \ref{sec:ANNDetection}, the ANND algorithm is presented. Simulation and performance results  are presented and discussed in section \ref{sec:Results}. Section \ref{sec:Conclusion} concludes the paper.

\section{MIMO-OFDM system model}\label{sec:SystemModel}
By combining multiple-input multiple-output and orthogonal frequency-division multiplexing technologies, wireless systems succeeded in obtaining high data rates and high spectral efficiency, which are not attainable for conventional SISO systems. These data rate and spectral efficiency enhancements with the MIMO and OFDM schemes stem from the parallel transmission technologies they offer in the space and frequency domains \cite{ciminiofdm1985}. Another benefit of combining OFDM with MIMO  is their joint robustness to frequensy-selective fading channels. 

\begin{figure}[ht]
    \centering
    \includegraphics[width=0.65\linewidth]{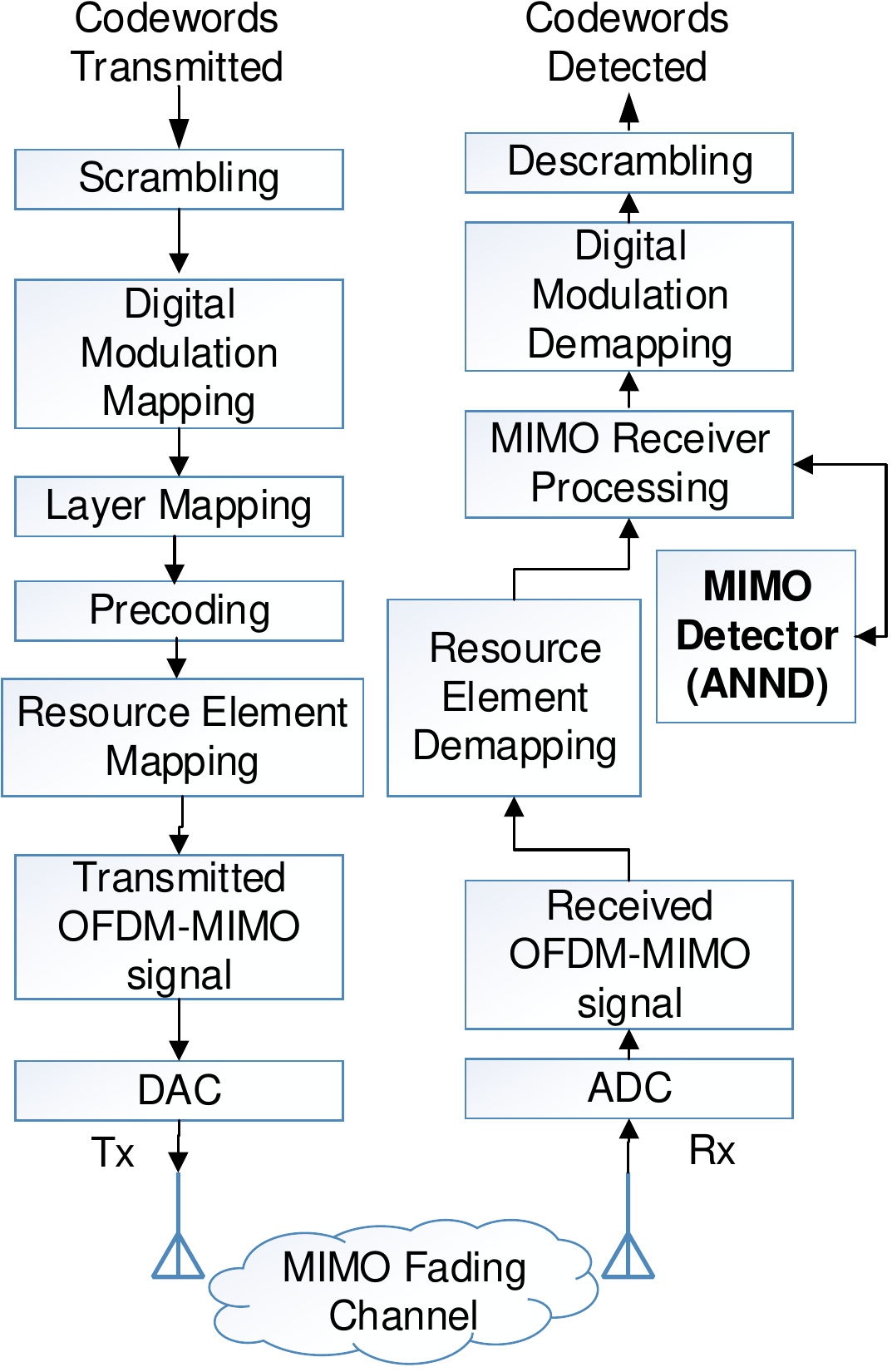}   
    \vspace*{-0.0in} 
\caption{ MIMO-OFDM system model.}
    \label{fig:PhyLayer}
\end{figure}

We consider here a MIMO-OFDM system with $N$ transmit antennas and $M$ receive antennas---thus having an $M\times N$ channel matrix and $N_s$ subcarriers. The structure of the transceiver is shown in Fig. \ref{fig:PhyLayer}.
The transmitted signal is formed into OFDM symbols by applying the inverse fast Fourier transform as follows
\begin{equation}
x_n(l)  = \sum\limits_{k=0}^{N_s-1}X_n(k) e^{j2\pi kl/N_s} , \;  l = 0, 1,\dots, N_s - 1,
\end{equation}%
where $X_n(k)$ is the  transmitted signal of $n^{\text{th}}$ transmit antenna on $k^{\text{th}}$ subcarrier. Also, to avoid inter-symbol interference (ISI), a cyclic prefix (CP) is inserted. After CP insertion, OFDM signals are ready to be transmitted. At the receiver, just after the ADC, fast Fourier transform (FFT) operates as a way of OFDM demodulation. Thus the input-output relation of the MIMO-OFDM system model can be expressed in a matrix form as follows
\begin{equation}
\mathbf{y} =\mathbf{H}\mathbf{x}+\mathbf{n},
\end{equation}%
where $\mathbf{y}=[y_1,...,y_M]^T$ and $\mathbf{x}=[x_1,...,x_N]^T,$ are the received and transmit signal vectors; $\mathbf{H}$ is the $M\times N$ channel matrix; $\mathbf{n}$ is the $M\times 1$ additive white Gaussian noise (AWGN) vector whose elements are zero mean and with variance of $\mathbf{\sigma}^2$; and $T$ denotes transpose operation.

After ADC, the FFT operates on the received OFDM symbols and the symbol on the $m^{\text{th}}$ antenna over the $k^{\text{th}}$ sub-carrier with $k = 0, 1,..., N_s - 1$ is written as
\begin{equation}
Y_m(k)  =\frac{1}{N_s}\sum\limits_{l=0}^{N_s-1}y_m(l) e^{-j2\pi kl/N_s}.
\end{equation}%
Hereafter, if conventional detectors were deployed, the pilot symbols could be extracted. However, detection with these methods is not very effective at low-SNR values or poor CSIR, which means applying those algorithms is not an optimum choice. Thus, we seek a new  method to achieve better detection performance. We propose a new ANN-based detector based on classification technique with the Levenberg-Marquardt technique\cite{demuth2014neural}.

\section{Artificial Neural Network}
\label{sec:MLP}
 An artificial neural network is a computing system which is inspired by a biological neural network. An ANN is formed of a collection of units called artificial neurons inter-connected, by synaptic weights. Several classes exist in terms of neurons connection types, activation functions, and learning approaches. In this paper we adopt the multilayered perceptron neural networks (MLP) structure \cite{demuth2014neural} and the Levenberg-Marquardt learning method.

\subsection{Multilayered Perceptron Neural Networks}
\begin{figure}[ht]
    \centering
    \includegraphics[width=0.9\linewidth]{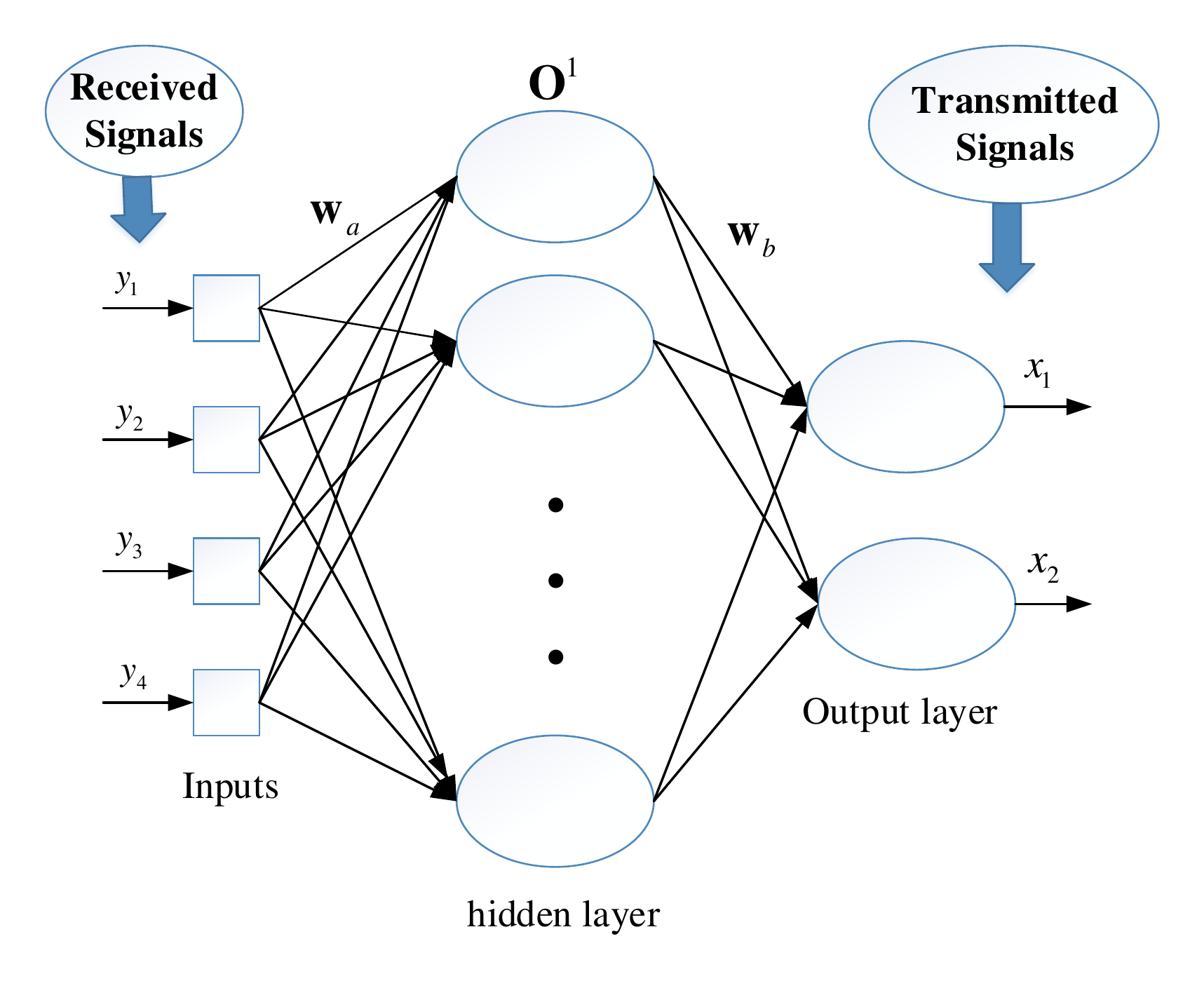} 
    \vspace*{-0.1in}
    \caption{ MLP neural network .}
    \label{figure:ANN3}
\end{figure}

Based on the type of connection between the neurons, several types of ANN are defined. The most common type of ANN is feed-forward multilayered perceptron (FF-MLP). An MLP consists of several layers which have, at least, an input layer, a hidden layer, and an output layer. Each node is a neuron that uses a nonlinear activation function
\cite{haganmlp1994}. In an FF-MLP, the input signal is passed through an activation function to produce
the output of the neuron. There can be hidden neurons that have an internal role in the network \cite{demuth2014neural}. Fig. \ref{figure:ANN3} shows a simple model of an ANN which consists of input, hidden, and output layers.
Once a neural network is trained for  a specific task, its weights can be adopted by another ANN for a similar task.

\subsection{Levenberg-Marquardt Algorithm}
The Levenberg-Marquardt (LM) algorithm is used to solve non-linear least squares optimization problems arising from the training phase. In order to adjust the weights of the hidden and output layers, we define an error function $\mathbf{E(w)}$ and minimize it during the training phase as follows
\begin{equation}\label{eq:errorfunc}
E(w) =\dfrac{1}{2}\sum\limits_{r=1}^{R}(d_r - x_r)^2,
\end{equation}%
where $d_r $ is $r^{\text{th}}$ desired output, $x_r$ is $r^{\text{th}}$ actual output and $R$ is the number of output points. At each iteration, the weights should be updated according to 
\begin{eqn}\label{eq:update}
\Delta \mathbf{w} =-(\mathbf{J}^T \mathbf{J}+\phi \mathbf{I})^{-1}\mathbf{J}^T E,
\end{eqn}%
where $\Delta \mathbf{w}$ is the weight difference vector, $\mathbf{J}$ is the Jacobian matrix containing the derivatives of the network errors with respect to the weights and biases, $\phi$ is the learning rate which determines the rate at which the weights $\mathbf{w}$ are updated at each step. Choosing a high learning rate leads to faster training, at the price, however, of having longer convergence time. So there is a trade-off between the learning and convergence rates. While a too large value
for $\mathbf{\phi}$ accelerates the training process, it may cause oscillation, can prevent the algorithm from reaching converge. On the other hand, a too small learning rate causes the algorithm to take a long time to converge. We can wisely optimize it during the learning phase based on the time taken by our network to converge.

\section{ANN-Based Signal Detection}\label{sec:ANNDetection}

In this section, we propose a blind detection technique for our MIMO-OFDM system model based on the classification approach in ANNs.

\subsection{ANN-Based Detector}
Our proposed ANND structure has two layers: one hidden and one output layer. We consider a tangent hyperbolic function as the activation function in each neuron of the hidden layer with seven neurons and hard-limit activation function in the output layer. The neural structure with four inputs and two outputs is used to adapt  OFDM signals to the neural network. The received symbols consist of complex signals whereas the neural network uses real signals. In order to adapt the neural network to the system, each complex signal is separated into real and imaginary parts. The mathematical operations in the layers of the network during the working and training phases are as follows: 
\begin{eqn}
net^1_j& =\sum\limits_{i=1}^{l_1}(y_i . w_{aji}),   \quad  j = 1, 2,\dots, 7,  \\
o^1_j &=f(net^1_j)=
{(e^{2net^1_j}-1)/(e^{2net^1_j}+1)},\\
net^2_r &=\sum\limits_{j=1}^{l_2}(o^1_j . w_{brj}), \quad  r = 1, 2,  \\
x_r&=f(net^2_r)=net^2_r,
\end{eqn}%
where $w_{aji}$  is the weight of the hidden layer's input at $j^{\text{th}}$ node, $l_1$ is the number of input nodes set here to 4, $w_{brj}$ is the weight of hidden to output layer at $r^{\text{th}}$  node,  and $l_2$ is the number of hidden-layer neurons set to 7 here.

\subsection{ANN Training Phase}
\begin{figure}[ht]
\hspace*{-0.0in}
    \centering
    \includegraphics[width=0.8\linewidth]{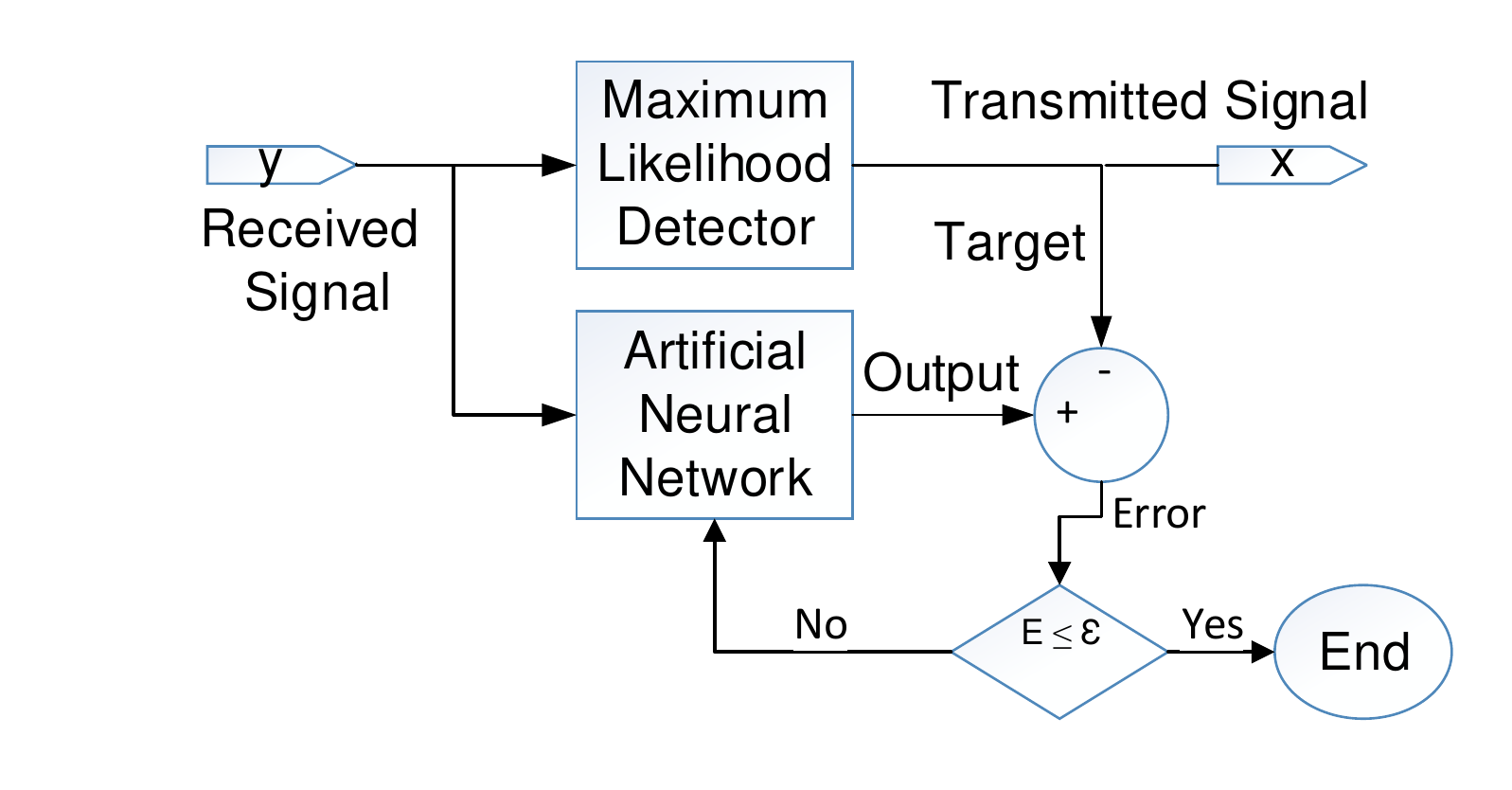}
    \vspace*{-0.0in}
    \caption{Training algorithm of ANN }
    \label{fig:ANN1}
\end{figure}

The training algorithm for a feed-forward neural network is as shown below:

\begin{itemize}
\item[1)] Initialize the weights vector $\mathbf{w}$ and learning rate $\phi$;
\item[2)] Take the received signal as an input and estimate the transmitted signal using ML detector as a target according to Fig. \ref{fig:ANN1};
\item[3)] Compute error function as per (\ref{eq:errorfunc});
\item[4)] Compute the weight difference $\Delta \mathbf{w}$ as per (\ref{eq:update});
\item[5)] Recompute the error function after using new weights according to $\mathbf{w}+\Delta \mathbf{w}$; 
\item[6)] If the error is smaller than the one computed in Step 3, then reduce the learning rate ($\mathbf{\phi}$) by 0.1 times; otherwise, increase it 2 times and go to Step 2;
\item[7)]Finish the training phase if the error function is less than the predefined value.
\end{itemize}

\section{Results and Discussion}\label{sec:Results}
\begin{figure}[ht]
\vspace*{0cm}
\centering
\subfloat[$(N_t, N_r) = (2, 2)$]{\hspace*{0cm}
\includegraphics[width=0.8\linewidth]{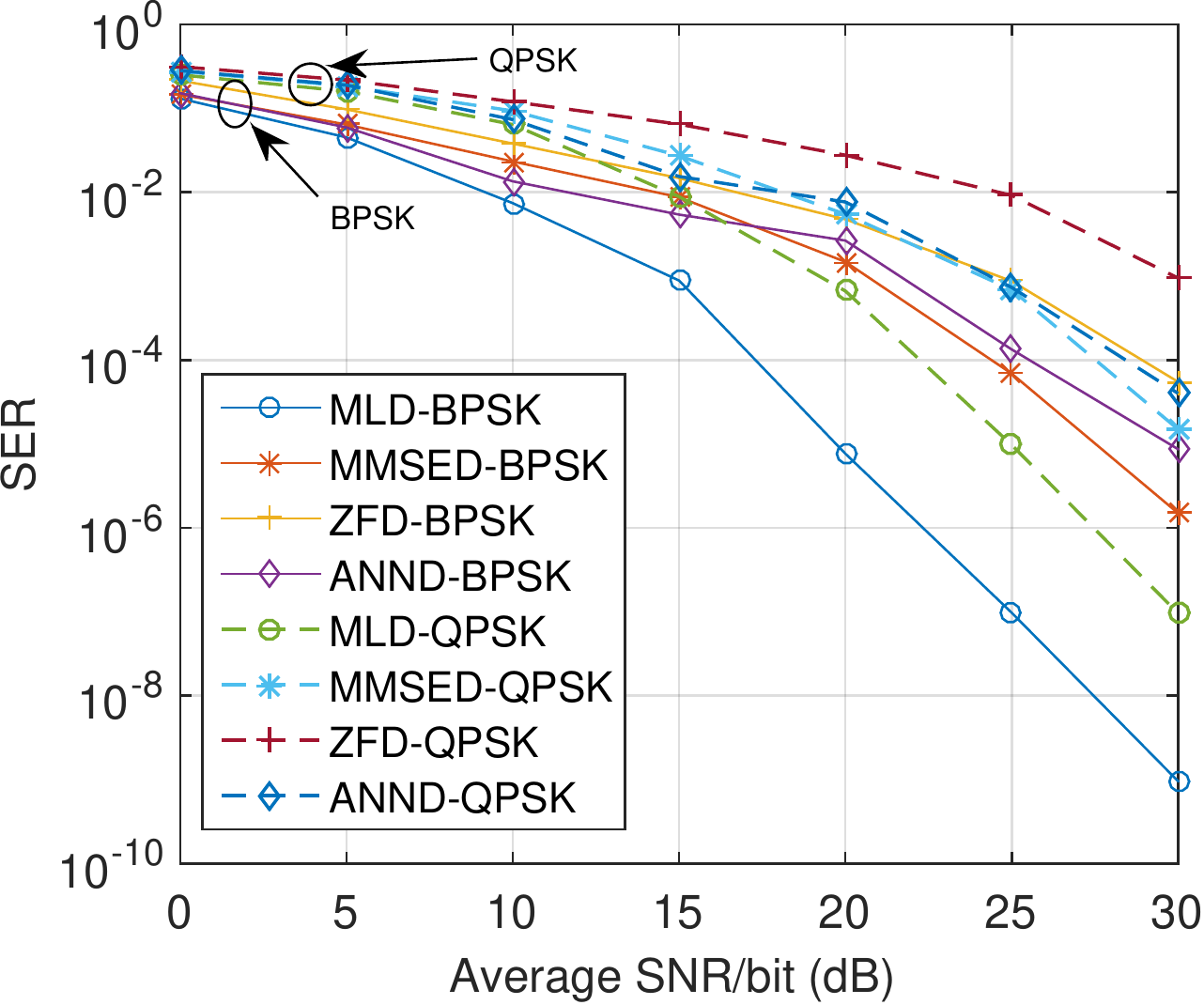}\label{fig:BQPSK-a}}
\vspace*{0cm}
\subfloat[$(N_t, N_r) = (4, 4)$]{\hspace*{0cm}
\includegraphics[width=0.8\linewidth]{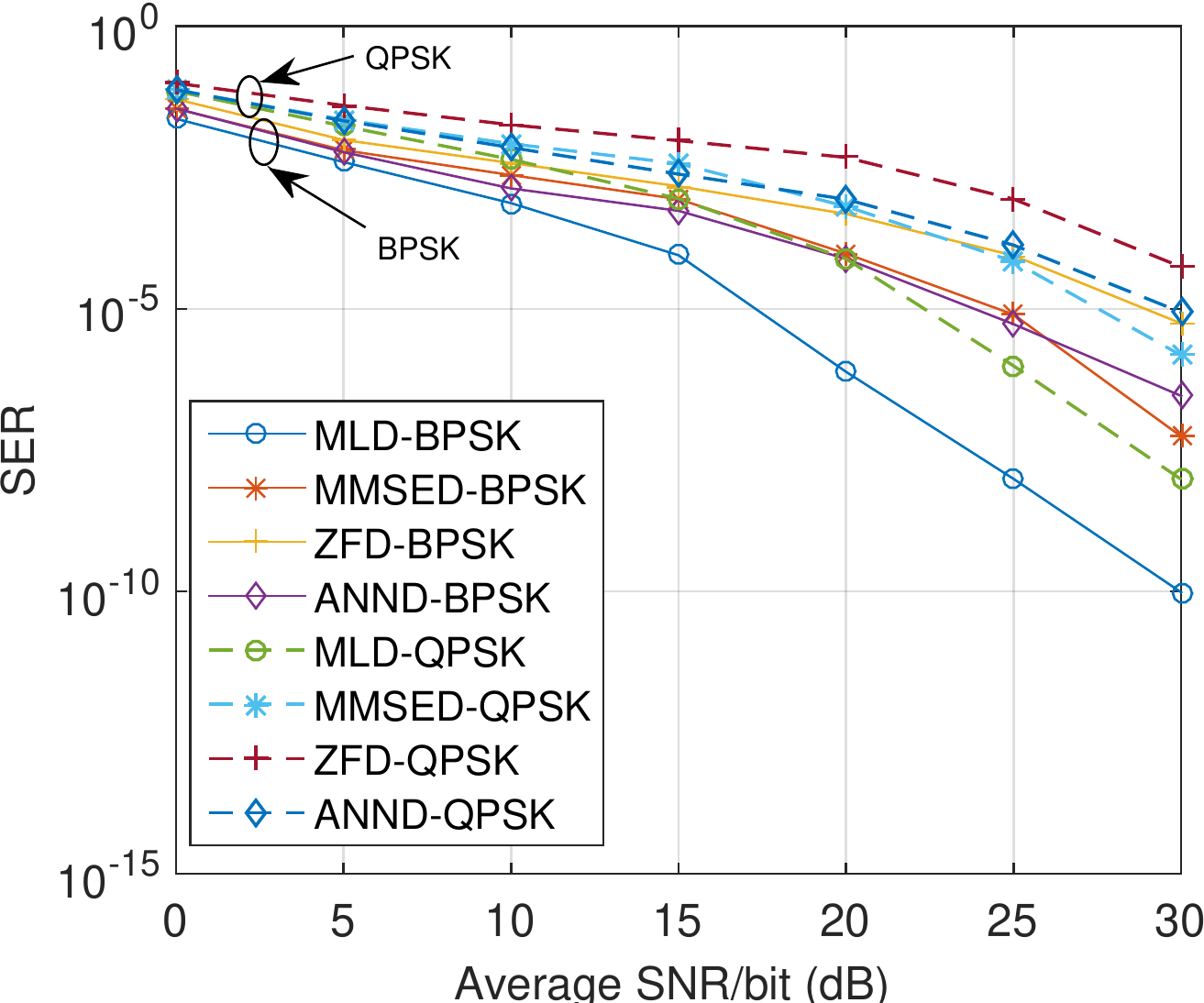}\label{fig:BQPSK-b}}
\vspace*{0cm}
\caption{SER performance of the proposed and conventional detection techniques for MIMO-OFDM systems for BPSK and QPSK.}\label{fig:BQPSK}
\vspace*{-0.0cm}
\end{figure}

\begin{figure}[!t]
\vspace*{0cm}
\centering
\subfloat[$(N_t, N_r) = (2, 2)$]{\hspace*{0cm}
\includegraphics[width=0.8\linewidth]{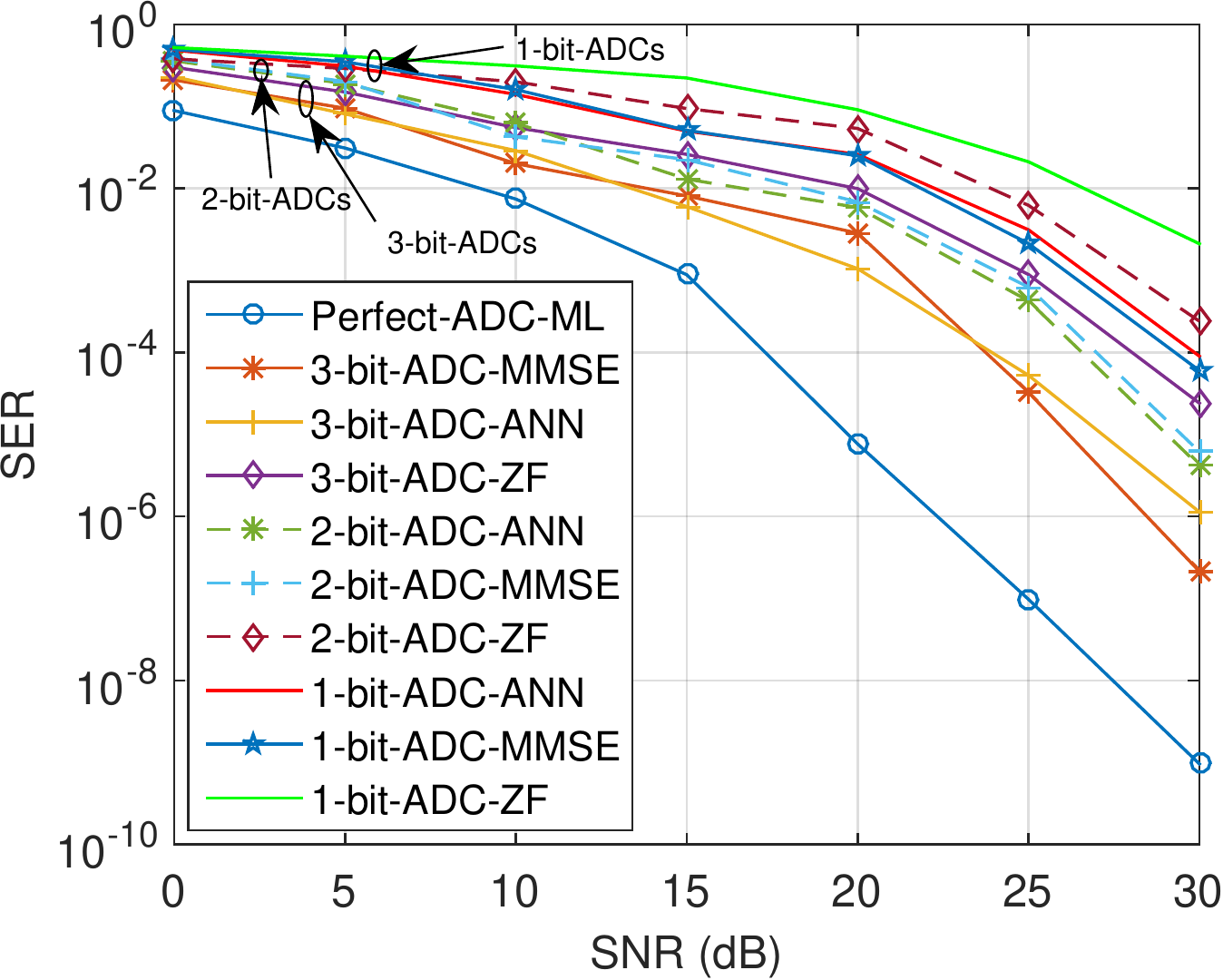}\label{fig:QPSKADC-a}}
\vspace*{0cm}
\subfloat[$(N_t, N_r) = (4, 4)$]{\hspace*{0cm}\includegraphics[width=0.8\linewidth]{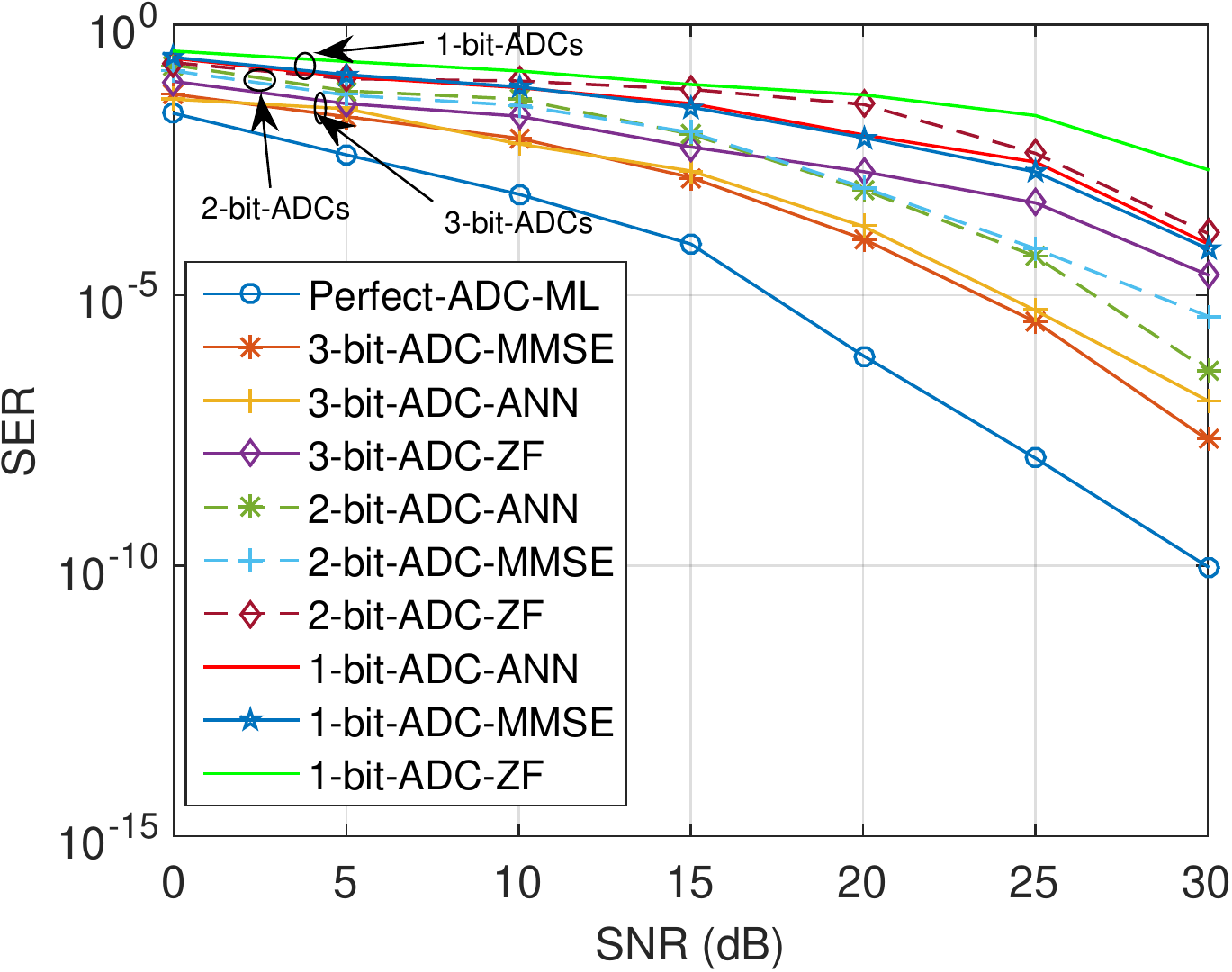}\label{fig:QPSKADC-b}}
\caption{SER performance of the proposed and conventional detection techniques for various numbers of ADC bits for MIMO-OFDM systems for QPSK.}\label{fig:QPSKADC}
\end{figure}

In this section, based on the extensive simulations, we compare the performance of our proposed ANN-based detector (ANND), in terms of SER, with the conventional ML, MMSE, and ZF detection techniques over a MIMO-OFDM system. We do such comparison versus the SNR as shown in Figs. \ref{fig:BQPSK} and \ref{fig:QPSKADC}. We compare in terms of computational time the complexities of these detectors in Table \ref{table_III}.
The simulation parameters of our MIMO-OFDM system and the multilayered percepteron neural network are listed in Tables \ref{table_I} and  \ref{table_II}. 
\begin{table}[ht]
\renewcommand{\arraystretch}{1.3}
\caption{MIMO-OFDM system parameters. }
\label{table_I}
\centering
\begin{tabular}{c||c}
\hline
\bfseries Parameter & \bfseries Value\\
\hline\hline
Carrier frequency $f_c$ & 5 GHz\\
Sampling frequency $f_s$ & 3 MHz\\
FFT size & 64\\
Modulation type  &  BPSK, QPSK\\
Channel type & flat Rayleigh channel\\
Number of antennas & $2\times2$ , $4\times4$\\
SNR Range & 0--30 dB \\
MIMO receiver equalizer & ML , MMSE , ZF, ANND\\
ADCs &  perfect and imperfect
\end{tabular}
\end{table}
\begin{table}[ht]
\renewcommand{\arraystretch}{1.3}
\caption{ANN Parameters.  }
\label{table_II}
\centering
\begin{tabular}{c||c}
\hline
\bfseries Parameter & \bfseries Value\\
\hline\hline
Number of inputs & 4 \\
Number of outputs & 2\\
Number of hidden
layers & 1\\
Number of neurons in
hidden layer & 7\\
Epoch number /
Iteration  &  1000\\
Training function /
Algorithm &  Levenberg-Marquardt \\
Performance metric & Mean square error\\
Target
error function $E$ & $10^{-3}$\\
Initial learning rate $\mathbf{\phi}$ &  0.35
\end{tabular}
\end{table}

we compare in Fig. \ref{fig:BQPSK} the SER performance  of the ANND, ML, MMSE, and ZF detection techniques using binary-PSK (BPSK) and quadrature-PSK (QPSK) modulation schemes when a perfect ADC (i.e. with infinite resolution) is deployed in the receiver. Fig. \ref{fig:BQPSK-a} and Fig. \ref{fig:BQPSK-b} are related to a $2\times 2$ and $4\times 4$ MIMO system, respectively. In  Fig. \ref{fig:BQPSK-a}, ANND outperforms ZF at any SNR value for both BPSK and QPSK Without, however, surpassing the optimal detector ML. In comparison to MMSE, the superiority of ANND depends both on the the modulation scheme and the SNR value. While for BPSK, ANND outperforms MMSE at low-SNR values, it performs worse with QPSK at high-SNR values. The difference in performance is due to limited training time and number of hidden layers.

Figs. \ref{fig:QPSKADC-a} and \ref{fig:QPSKADC-b} illustrate the SER  achieved by the proposed ANND, MMSE and ZF using different LR-ADC resolutions against the SER obtained by ML with perfect ADC taken as a lower band, when employing QPSK modulation over  $2\times2$ and $4\times4$  antennas size.It is observed that the proposed detection technique provides a  SER reduction in system  compare to MMSE algorithm in all various numbers of ADC bits, specially in case of $4\times4$  antennas size.
From this it can be stated that the neural detector has better performance than MMSE and ZF algorithms for MIMO-OFDM systems with low-resolution ADCs, especially in low SNR range, with not only $4\times4$ MIMO structures, respectively. Obviously, the proposed ANND outperforms both ZF and MMSE, more so at lower SNR values, lower ADC resolutions, and with a larger $4\times4$ MIMO structure. Besides, as can be seen from Table \ref{table_III}, the proposed ANND has a lower complexity than ML in terms of computational time---a net advantage that further increases with large-scale MIMO structures such as massive MIMO
systems.
\begin{table}[!h]
\renewcommand{\arraystretch}{1.3}
\caption{THE COMPUTATIONAL TIME COMPLEXITY }
\label{table_III}
\centering
\begin{tabular}{c||c}
\hline
\bfseries Algorithms & \bfseries  Time complexity (/ML)\\
\hline\hline
MLD & 1.000\\
MMSED & 0.0511\\
ZFD & 0.0425\\
ANND  &  0.3429\\
\end{tabular}
\end{table}
\section{Conclusion}\label{sec:Conclusion}
We proposed a new signal detection method based on artificial neural network. We considered a MIMO-OFDM system with perfect and imperfect analog-to-digital converters. We compared the proposed ANN-based detector with the conventional of maximum-likelihood, minimum mean square error, and zero-forcing detection methods in both ADC cases with two different modulation schemes. Extensive simulations suggest that ANND offers a very promising alternative for applications requiring large numbers of antennas, such as massive MIMO.

\bibliographystyle{IEEEtran}


\end{document}